\documentstyle [epsf,multicol,aps] {revtex}

\newcommand{\beq}{\begin{equation}}
\newcommand{\eeq}{\end{equation}}

\draft
\title{ Barkhausen-like conductance noise in polycrystalline high
Tc superconductors immersed in a slowly varying magnetic field.}
\author{P.Mazzetti, A.Stepanescu and P.Tura}
\address{Istituto Nazionale per la Fisica della Materia, Unit\`a di Ricerca del Politecnico
di Torino, and Dipartimento di Fisica, Politecnico di Torino,Corso
Duca degli Abruzzi 24, I-10129 Torino, Italy}

\author{A.Masoero}
\address{Istituto Nazionale per la Fisica della Materia, Unit\`a di Ricerca del Politecnico
di Torino, and Dipartimento di Scienze e Tecnologie Avanzate,
Universit\`a del Piemonte Orientale 'Amedeo Avogadro' \\ C.so
Borsalino 54, I-10131 Alessandria, Italy}

\author{I.Puica}
\address{Catedra de Fizica, Universitatea Politehnica Bucuresti, Spl. Independentei 313,
Bucharest, Romania}

\date{\today}

\begin{document}
\maketitle

\begin{abstract}

Analysis of the resistive transition of a polycristalline YBCO
specimen produced by an a.c. magnetic field, reveals the presence
of a large conductance noise signal which is repetitive over
subsequent magnetization cycles. It is shown that the noise arises
from avalanche effects produced by the simultaneous resistive
transition of large groups of weak links. Owing to its
repeatability, the noise signal may be considered a sort of
signature of the weak links critical current distribution, making
it an interesting new tool for the study of high Tc ceramic
superconductors, as reported measurements of noise hysteresis
show.

\end {abstract}
\begin{multicols} {2}
The resistive transition of polycrystalline high Tc
superconductors is, in general, dominated by the intergrain
layers, which behave as shunted Josephson junctions (weak links).
The weak link ensemble, on the other side, is characterized by a
distribution of critical currents which depends on the local value
of the magnetic field. When the specimen is crossed by a d.c.
current and immersed in a slowly increasing magnetic field, there
are gradual transitions to the resistive state of the weak links
and, above a given threshold, the whole specimen becomes
increasingly resistive. As shown by the reported experiments, this
process is not smooth, but is characterized by a large noise. It
is very interesting to notice that the noise signal represents a
sort of signature of the local distribution $\cite{note}$ of the
weak-link critical currents, since it is \textit{repetitive} over
different magnetic field cycles, until a magnetic perturbation,
strong enough to change such a distribution by effect of flux
trapping within the superconducting grains, is applied.\\ Another
important aspect of this noise, which has a completely different
origin of the noise observed in stationary conditions $[2-6]$,
concerns the fact that it is produced by strong correlations
during the transition of the weak-links, giving rise to avalanche
effects, similar to those observed in the Barkhausen noise of
ferromagnets. As computer simulations show, these avalanches
correspond to abrupt rearrangements of the internal current
distribution, and thus of the distribution of superconducting and
resistive weak-links, due to the need of satisfying Kirchoff
equations in a network of highly non linear circuit elements. The
step-like variation of the specimen conductance in correspondence
to each weak-link transition avalanche gives rise, as reported in
the following, to a $1/f^2$ power spectrum, as opposed to the
$1/f$ spectrum observed in the stationary case $\cite{step}$. Also
the amplitude of the noise, which is orders of magnitude larger
than expected in the case of uncorrelated transition of the
individual weak links, confirms the presence of these avalanche
effects.\\ Another important aspect, also confirmed by computer
simulations, concerns the fact that the position along the time
axis and the amplitude of these steps, when the magnetic field is
varied, is strongly dependent on the local distribution of the
weak-link critical currents. Thus the detection and storing of the
noise signal within a given field interval along the loop allows
to observe very small variations of such a distribution, produced
by external perturbations. For instance, as shown in the
following, small hysteretic effects, invisible on the conductance
vs. field plot, may be evidenced by comparing the noise signals
detected in the same field interval along a loop when the field is
increasing or decreasing. It should be noticed that the
repeatability of the noise signal during cycling is a very rare
property of a physical system, and reflects an intrinsic
non-fluctuating aspect of the nature of the system itself. In the
case of the Barkhausen noise, for instance, the noise signal is in
general not repetitive over different cycles, since it is
determined during each cycle by the domain
configuration obtained in the preceding one $\cite{Durin}$.\\
Experiments were performed on a YBCO specimen suitably treated to
lower the weak-links critical currents. This treatment made the
specimen very sensitive to low magnetic fields and allowed to use
small, noise free a.c. magnetic fields to perform the experiments.
The use of small fields was also needed to avoid flux penetration
within grains, such that only the weak-links were involved in the
resistive transition of the specimen. Noise was detected by using
a 4 contact technique on a small cylindrical specimen 1mm in
diameter and 10mm in length. An extra low noise input transformer
was used to match the input impedance of the preamplifier and to
cut the d.c. component of the noise signal. It acted as a
band-pass filter in the range between 5Hz and 10kHz. A standard
electronic set-up was used to digitally
storing  the signal in a selected field interval during cycling.\\
Fig.1 shows three samples of the noise signal detected within the
same field interval when looping and the background noise when the
field is kept constant. The repeatability of the noise signal
before the application of the magnetic perturbation is quite
evident. Small changes are mainly due to the superimposed
background noise.\\ In Fig.2 the power spectra of the noise
obtained during field variation and in a steady state condition
are reported. The near $1/f^2$ behaviour of the spectrum in the
first case is consistent with the assumption that the noise is
constituted by the superposition of random step-like variation of
the specimen conductance. The cut-off frequencies of the input
transformer  had the effect of limiting the frequency range over
which the noise spectrum could be detected. They had also the
effect of transforming the voltage steps into exponential pulses,
a fact that explains why the noise signals looks like those
reported in Fig.1.\\ These results suggest that the presence of
abrupt discontinuous conductance changes in a network of non
linear resistors, having current-voltage characteristics similar
to the ones of Josephson junctions, be a general property of this
type of networks. This fact is confirmed by computer simulations,
as reported in the following. Below the critical temperature of
the grains, the polycrystalline superconductive sample is modelled
as a rectangular three-dimensional array of intergranular weak
links, fed with constant d.c. current flowing between two plane
electrodes applied on the opposite faces. Any individual weak link
is viewed as a nonlinear circuit element which is in a
superconducting state for local currents below a certain critical
value \textit{i}$_{c}$ and in a normal resistive state for
currents above this value (see Fig.3). Its critical current value
depends in turn on the temperature and on the local intergranular
magnetic field B$_{i}$. We used standard relations $[9-12]$ to
determine the dependance of the weak-link critical current
\textit{i}$_{c}$ on the local field value B$_{i}$, which, owing to
the screening effect of the superconducting grains, is always
tangent to the weak-link surface and depends on the angle $\theta$
between the applied field {\bf B} and the normal {\bf n} to the
weak-link surface. In the calculations, whose details will be
given in another paper, the relation between the local
intergranular field B$_{i}$ and the applied field B was assumed to
be of the form \beq B_{i}={K}\cdot{B}|\sin\theta| \eeq where K is
a flux compression factor due to the flux exclusion within the
superconducting grains. The large spread of the values of $\theta$
gives rise to a large distribution of critical currents
\textit{i}$_{c}$, whose average value depends on the modulus of
the applied field B. In the calculations the values of the local
angle $\theta$ were set by cosidering an uniform solid angle
distribution for the junction orientations.\\ Kirchoff's equations
for potentials and currents, together with the weak-link
current-voltage characteristic, provided a set of coupled
nonlinear equations for the electric potentials at the nodes of
the network, which were solved by an iterative method. At each
iterative step, from the previous set of node potentials, the
junction conductances were actualized according to the nonlinear
current-voltage characteristic described above. Considering then
these conductances as constant, the linear equation system given
by Kirchoff's law is solved, so that a new set of node potential
is found. The calculation technique, a modified algorithm based on
the method of the conductance matrix, is described in ref.
$\cite{Cattaneo}$. The iterative cycle ends when the difference
between two consecutive sets of node potentials becomes
sufficiently small with respect to their actual values. We used in
our simulations a relative tolerance of the order of 10$^{-7}$.
Results corresponding to two different actualization of the same
distribution of junction critical currents in a cubic network of
about 1000 junctions are reported in Fig.3. Several checks were
made to ensure that the steps were not an artifact of the
numerical calculations by increasing the point density and by
reducing the relative tolerance for exiting the iteration loop.
For what concerns this tolerance, it was observed that the
position of the steps depends on this quantity when its
 value is not low enough, but that their position and amplitude becomes stable when it
is assumed to be sufficiently small ($<$10$^{-6}$). It was also
observed that the position of the steps along the field axis is
very sensitive to small changes in the assumed distribution of the
critical currents, a fact that is typical of the presence of
instabilities in a complex system. The high sensitivity of the
noise signal to even small variations of the local distribution of
the weak link critical currents allows to use this quantity as a
test for evidencing effects that would not be detectable by other
means. We present here some experimental results that clearly show
the presence of magnetic hysteresis in the YBCO specimen described
above, even when it is submitted to an a.c. magnetic field whose
amplitude is so small that hysteresis cannot be detected by
conventional means.\\ In this experiment the specimen was
submitted to an a.c. field of a few Gauss of amplitude and crossed
by a d.c. current of a few mA. Fig.4 represents the behaviour of
the electrical resistance of the specimen vs. magnetic field. It
appears to be completely reversible and without evident
hysteresis. Fig.5 shows the noise signal detected in a small field
interval during cycling. It is evident that the noise signal,
which is repetitive over subsequent cycles, is quite different
along the increasing or decreasing branch of the loop, even if the
field interval where the signal is detected is the same. This fact
proves the presence of hysteresis, probably due to some magnetic
flux trapping along the cycle. This effect could also be related
to an intrinsic hysteresis of the v-i characteristic of each
individual weak link, a possibility which, however, was not taken
into account in the computer simulations. In this case it was
tested that a complete reversibility is obtained. From the
experimental point of view, it should however be possible to
distinguish between flux trapping and hysteresis of the individual
weak link characteristics by performing experiments similar to the
reported ones, but progressively reducing the field amplitude to
see if noise tends to become reversible. It may be noticed that
measurements performed on what could be considered a single weak
link, i.e. the grain boundary junction of a bicrystal of BSCCO
$\cite{Marx2}$ and a submicron grain boundary junction in YBCO
thin films $\cite{Herbstritt}$ show Josephson like characteristics
but no hysteretic effects were reported by the authors. In any
case it seems clear that the noise could be used to investigate
this and other aspects of the resistive transition of granular
superconducting materials. A more complete analysis of the effect
on the noise signal of field amplitude, frequency, temperature and
d.c. current will be the argument of an extended paper. Here we
want to stress that the noise signal, being an highly structured
quantity, allows to evidence even very small changes of the weak
link critical current distribution. Its repeatability over
different magnetization cycles allows to depurate it from the
background noise through a simple averaging procedure, making it
an interesting new tool to analyze the physical properties of
polycrystalline HTCS. Measurements on different materials, either
bulk or thin films, might reveal important aspects of the physics
of these superconductors, related to their granular structure. At
the moment no checks were made to see what happens in
monocrystalline bulk or thin films HTCS.

\begin{figure}[htb]
\narrowtext \centerline{
        \epsfxsize=7cm
                \epsfbox{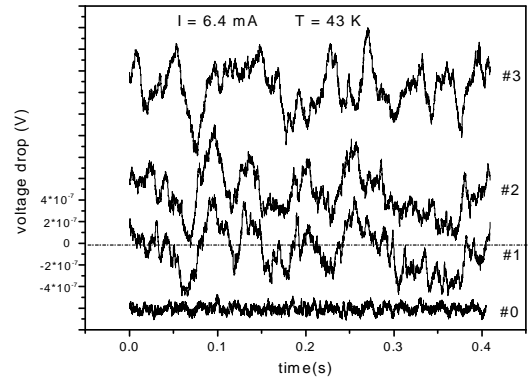}
                    }
\vspace*{0.5cm} \protect \caption {Voltage noise measured on the
HTCS specimen crossed by a current of 6.4 mA and immersed in an
a.c. triangular magnetic field of amplitude 3.2 G and frequency
0.08 Hz. Curves $\sharp$1, $\sharp$2, $\sharp$3 show the noise
taken on the same field interval between 0$\div$0.4 G over
different cycles. Curve $\sharp$0 shows the background noise when
the field is kept constant. The similarity of curves $\sharp$1,
$\sharp$2 shows that roughly the same processes giving rise to the
noise occur at the same values of the field when no large magnetic
perturbations are applied to the specimen. Curve $\sharp$3 shows
the effect of the momentary application of a magnetic field of a
few tens of Gauss. The variation of the noise signal is attributed
to a permanent change of the critical current distribution caused
by flux trapping within the superconducting grains. } \label{fig1}
\end{figure}

\begin{figure}[htb]
\narrowtext\centerline{\epsfxsize=6.8cm \epsfbox{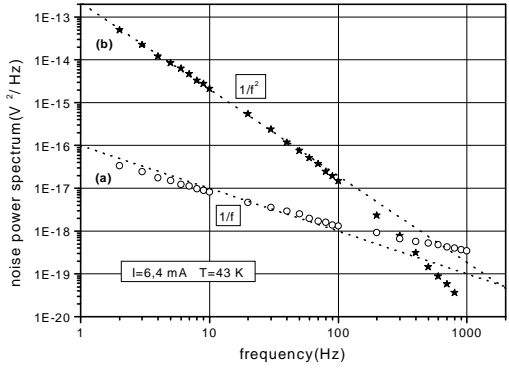}}
 \vspace*{0.5cm}
\protect\caption {Power spectra of the noise generated by an a.c.
magnetic field of 3.2 G amplitude and 0.08 Hz frequency (curve b)
and of the noise obtained when the field is kept constant (curve
a). Curve (a) has been subtracted from the spectrum obtained in
the presence of the a.c. field to obtain curve (b), which thus
represents the pure contribution to the noise of the a.c. field.
The nearly $1/f^2$ slope of this curve is typical of the spectrum
of a noise constituted by a series of randomly distributed steps.
Dotted lines represent the $1/f$ and $1/f^2$ slopes. Below 5 Hz
the data have been corrected by using the cut-off frequency curve
of the input transformer.} \label{fig2}
\end{figure}

\begin{figure} [htb]
\begin{center}
\narrowtext\centerline{\epsfxsize=6.8cm \epsfbox{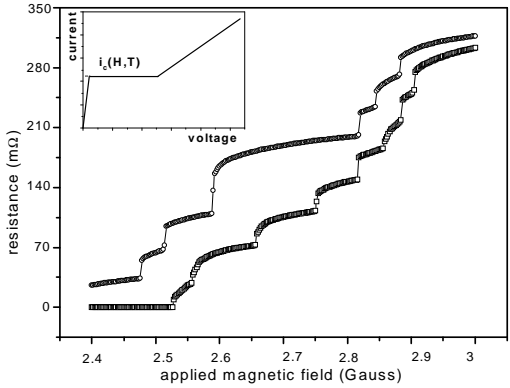}}
\end{center}
\vspace*{0.5cm} \caption {Results of a simulation of the behavior
of the electrical resistance of the YBCO specimen under a varying
magnetic field. The simulation refers to a 3-Dimensional cubic
network of about 1000 nonlinear resistor elements having a i-v
characteristic schematically represented in the inset. The two
curves are different simulations corresponding to different
actualization of the same distribution of critical currents
obtained by changing the sequence of the pseudo-random numbers
generated by the computer. For a comparison with the results
reported in Fig.1 and Fig.2, account should be taken that the
input transformer cut the d.c. component of the signal and changes
the steps into exponential pulses. }\label{fig3}
\end{figure}

\begin{figure} [tbp]
\narrowtext\centerline{\epsfxsize=6.8cm\epsfbox{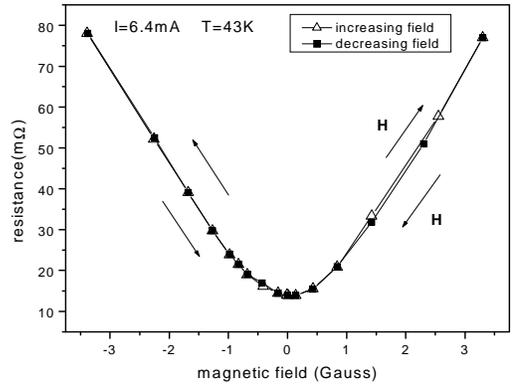}}
\vspace*{0.5cm} \protect\caption {Electrical resistance behaviour
vs. magnetic field, of the HTCS specimen crossed by a d.c. current
I slightly higher than its critical current at zero field. The
figure shows that the specimen resistance, in addition of being
very sensitive to the magnetic field, does not present any visible
hysteretic effect. However analysis of the noise signal allows to
evidence that some type of hysteresis as actually taken place. }
\label{fig4}
\end{figure}

\begin{figure} [tbp]
\narrowtext\centerline{\epsfxsize=6.8cm\epsfbox{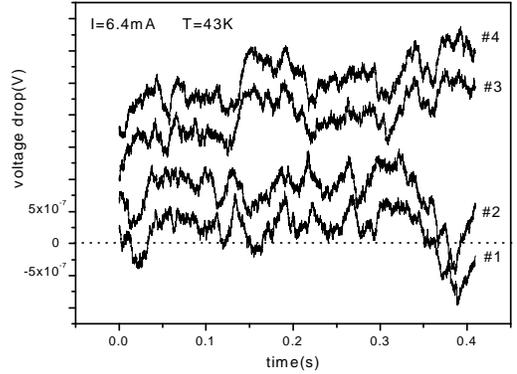}}
\vspace*{0.5cm} \protect\caption {Noise signals detected within
the same field interval of 0.4 G during cycling as in Fig.1.
Curves $\sharp$1 and $\sharp$2 are taken when the field is
increasing and curves $\sharp$3 and $\sharp$4 when the field is
decreasing. The change of the noise shape evidences the presence
of a magnetic hysteresis which changes the weak links critical
current distribution. This hysteresis is so small that it cannot
be evidenced by direct measurements (see Fig.4). } \label{fig5}
\end{figure}

\end{multicols}

\begin{thebibliography} {14}
\bibitem{note}
With "local distribution", as opposed to "statistical
distribution" we intend a particular actualization of the
statistical distribution over an ensemble of ordered elements.
\bibitem{step}
A $1/f^2$ power spectrum can be considered as the limit of a
lorentzian spectrum when its cut-off frequency tends to zero. It
is  the case of the power spectrum of a train of Poisson
distributed exponential pulses whose decay time tends to $\infty$
(steps).
\end{thebibliography}
\end{document}